\documentclass[12pt]{article}

\usepackage{amssymb, amsmath, xypic}
\usepackage{amsthm}
\usepackage{url}

\newtheorem{result}{Result}
\newtheorem{thm}{Theorem}
\newtheorem{remark}{Remark}

\usepackage{subfig,tikz}
\usetikzlibrary{decorations.pathmorphing}
\usepackage{graphics}

\DeclareMathOperator{\interior}{\mathsf{Int}}
\newcommand{\R}{\mathbb{R}}
\newcommand{\Z}{\mathbb{Z}}

\newcommand{\tr}{\,{\rm tr}\,}

\renewcommand{\d}[1]{\ensuremath{\operatorname{d}\!{#1}}}
\def\be{\begin{equation}}
\def\ee{\end{equation}}
\def\bea{\begin{eqnarray}}
\def\eea{\end{eqnarray}}

\title{ \bf{Notes on maximal slices of five-dimensional black holes}}

\author{Aghil Alaee\footnote{aak818@mun.ca },  Hari K. Kunduri\footnote{hkkunduri@mun.ca }  and Eduardo Mart\'inez Pedroza\footnote{emartinezped@mun.ca} \\ \\
\small \sl  Department of Mathematics and Statistics, \\  \small \sl Memorial University of Newfoundland \\ \small \sl St John's NL A1C 4P5, Canada}

\date{}

\begin{document}

\maketitle






\begin{abstract}
We consider maximal slices of the Myers-Perry black hole, the doubly spinning black ring, and the Black Saturn solution.  These slices are complete, asymptotically flat Riemannian manifolds with inner boundaries corresponding to black hole horizons. Although these spaces are simply connected as a consequence of topological censorship, they have non-trivial topology.  In this note we investigate  the question of whether the topology of spatial sections of the horizon uniquely determines the topology of the maximal slices. We show that the horizon determines the homological invariants of the slice under certain conditions.  The homological analysis is extended to black holes for which explicit geometries are not yet known.  We believe that these results could provide insights  in the context of proving existence of deformations of this initial data. For the topological slices of the doubly spinning black ring and the Black Saturn we compute the homotopy groups up to dimension $3$ and show that their $4$-dimensional homotopy group is not trivial.
\end{abstract}


\section{Introduction} Initial data for Einstein's equations are specified by $(\Sigma, h,K)$ where $\Sigma$ is a Riemannian manifold with metric tensor $h$ and second fundamental form $K$ regarded as a spacelike hypersurface in a spacetime.  Einstein's equations imply the constraints
\begin{eqnarray}\label{eq:constraints}
R_h - K^{ab}K_{ab} + (\tr K)^2 &=& 8\pi \mu \nonumber \\ \nabla^b\left(K_{ab} - \tr K h_{ab}\right) &=& -4\pi j
\end{eqnarray} where $(\mu,j)$ are the local energy density and momentum current respectively.  Sufficient conditions for the existence of complete, asymptotically Euclidean solutions to \eqref{eq:constraints}  are discussed in \cite{CIY} and with an additional apparent horizon boundary condition in \cite{Maxwell}. For time symmetric initial data $K_{ab}=0$, \eqref{eq:constraints} reduce simply to $R_h = 8\pi \mu \geq 0$ where $R$ is the scalar curvature of $(\Sigma,h)$.  In this special case, many important problems of general relativity translate into purely Riemannian ones, which is a useful simplification. For example, the Riemannian Penrose inequality establishes a sharp bound on the ADM mass $m$ \cite{Bartnik} of any asymptotically flat $(\Sigma,h)$ in three dimensions with $R_h >0$ in terms of the area of an outermost-area minimizing inner boundary \cite{Hu-Il,Bray}.  Such theorems are of significant mathematical interest, but they also clearly have important implications for general relativity.  In particular, a local form of the Riemannian Penrose inequality was recently used to study instabilities of higher-dimensional black holes \cite{FMR} and more generally, in connection to dynamical stability \cite{HW}.  \par 
An important class of hypersurfaces are \emph{maximal} slices, satisfying $\tr K = 0$.  Note that \eqref{eq:constraints} implies a necessary condition for their existence is $R_h \geq 0$.  Such slices play a useful role in studies of the initial value problem, singularities and numerical relativity (see, e.g. \cite{Andersson, CBY,Eardley}).  The existence of such slices for asymptotically flat spacetimes was proved under weak boundary conditions in Ref.\ \cite{Bartnik90}. It was proved that vacuum evolution of asymptotically flat data for the Einstein equations admit a foliation by `almost' maximal slices \cite{Bartnik90} (i.e. they are maximal outside a spatially compact set) and these results were subsequently strengthened and generalized \cite{ChruWald}. Furthermore, Witt has established topological obstructions to existence of maximal slices in globally hyperbolic spacetimes with both closed spatial and asymptotically flat spatial sections \cite{Witt1, Witt2}. In particular, these results demonstrate that asymptotically flat (globally hyperbolic) spacetimes admitting a maximal slice are rare\footnote{Since all $\Sigma$ occur as Cauchy slices of globally hyperbolic solutions to the Einstein equations \cite{Witt1,Witt2}}. \par
A consequence of topological censorship, which holds in all dimensions, is that the domain of outer communication of an asymptotically flat black hole must be simply connected.  Spacetimes admitting non-simply connected asymptotically flat Cauchy surfaces are singular by a higher-dimensional analogue of Gannon's theorem \cite{Gannon}.  In dimensions greater than 4, for globally hyperbolic spacetimes, Cauchy slices may exist which are simply connected but still have non-trivial topology. Such slices can be shown to admit asymptotically flat initial data satisfying the dominant energy condition \cite{Witt1, Witt2}. The recent singularity theorem \cite{SW} demonstrates that spacetimes admitting a (possibly simply-connected) Cauchy surface characterized by a certain nonpositive smooth invariant (a generalization of the Yamabe invariant for asymptotically flat manifolds) must collapse to form singularities. Hence for globally hyperbolic spacetimes, certain non-trivial topological structures must be hidden behind horizons; in particular, the Cauchy slices cannot have non-trivial Seiberg-Witten invariants \cite{SW}.

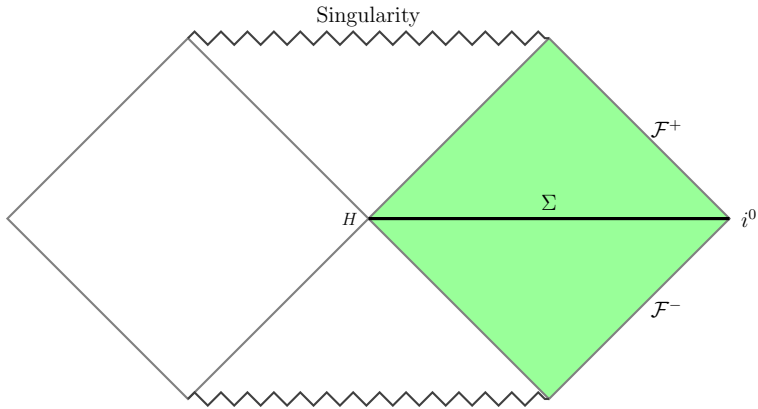
\begin{figure}[h]
\centering
\begin{tikzpicture}[scale=0.8, every node/.style={scale=0.7}]
\draw[gray,thick,fill=green!40!white] (0,0)node[black,at start,left=3pt]{$H$}--(3,3)--node[black,at end,right=2pt]{$i^0$}(6,0)node[black,midway,right=2pt]{$\mathcal{F^+}$}--(3,-3)node[black,midway,right=2pt]{$\mathcal{F^-}$}--(0,0);
\draw[gray,thick] (0,0)--(-3,3)--(-6,0)--(-3,-3)--(0,0);
\draw[gray!50!black,thick, decorate,decoration=zigzag] (-3,3) --node[black,above=2pt]{Singularity }(3,3);
\draw [gray!50!black,thick, decorate,decoration=zigzag]  (-3,-3) -- (3,-3);
\draw[black,very thick] (0,0)--node[black,above,midway]{$\Sigma$}(6,0);
\end{tikzpicture}
\caption{{\small The domain of outer communication has topology $\mathbb{R} \times \Sigma$.}}\label{fig:DOC}
\end{figure}

In four dimensions, the Kerr solution is the unique analytic, stationary asymptotically flat black hole vacuum solution.  The maximal analytic extension of the spacetime is of course not globally hyperbolic, but one can identify a particular useful class of initial data sets which are maximal.  In this case, $\Sigma$ can be shown to have topology $S^3 - \{0,\pi\}$ (an $S^3$ with the north and south poles deleted) \cite{DainPRL} . Spatial cross-sections of the event horizon, $H$, correspond to a minimal $S^2$ at the equator of this punctured $S^3$. If we restrict to the region consisting of one asymptotic flat end of $\Sigma$ with $H$ as its inner boundary, $(\Sigma,h)$ is a complete Riemannian manifold with boundary and represents a simply connected cobordism between $H \cong S^2$ and the `sphere at infinity' $S^2_\infty$ (see Figure \ref{fig:DOC}). This is, of course, required by topological censorship~\cite{topcen}, which asserts that all non-simply connected topological structures lie behind horizons.   

The situation in five dimensions is more complicated.  Consider stationary, non-static analytic asymptotically flat black hole solutions of the vacuum Einstein equations. The black hole topology theorem \cite{GS} asserts that provided the dominant energy condition holds, $H$ admits a metric of positive scalar curvature, and hence $H$ will be $S^3$ or one of its quotients, e.g. lens spaces, or $S^1\times S^2$, or connected sums thereof.  More recently it has been shown that if the spacetime admits an $\mathbb{R} \times U(1)$ isometry group\footnote{This must be the case for non-extreme, stationary rotating (analytic) black holes by the rigidity theorem \cite{HIW, IM}}, there are stronger restrictions not only on $H$ but the domain of outer communication \cite{HHI}.  In particular it has been proved that $H$ must either be a connected sum of \emph{at least} one $S^1 \times S^2$ and lens spaces or one of several possible quotients of $S^3$ by isometries. The latter class includes,  in addition to lens spaces,  the Poincare homology sphere and Prism manifolds \cite{HHI}.  Furthermore these authors proved
\begin{thm}[Hollands, Holland, Ishibashi]\cite{HHI} \label{theorem1} Consider an analytic, stationary, rotating vacuum black hole spacetime with isometry group $\mathbb{R} \times U(1)$. Then the domain of outer communication has topology $M \cong \mathbb{R} \times \Sigma$ where
\begin{equation}
\Sigma \cong \left(\mathbb{R}^4 \# n \, (S^2 \times S^2) \#n' \, (\pm \mathbb{CP}^2 )\right) - B
\end{equation}  where $B$ is a compact manifold with boundary $\partial B\cong H$ and $(n,n') \in \mathbb{Z}$.
\end{thm} We will only  focus on the case $n = n' = 0$ for the rest of this note. Note that if $M$ admits a spin structure then $n' = 0$.  Observe that $n=n'=0$ is a non-trivial restriction.  For example, one can consider asymptotically flat solutions for which $n \neq 0$ (corresponding to `bubbles' in the domain of outer communications).

We consider $\Sigma$ to be a simply connected manifold with boundary, $\Sigma \cup B \cong \mathbb{R}^4$ and $\Sigma$ and $B$ have a common boundary, $\Sigma \cap B \cong H$ . Following \cite{HHI}, we will refer to $B$ as the `black hole region'.  Since much of our analysis is at the topological level, we may study properties of $\Sigma$ arising as slices of more general black holes  for which explicit analytic solutions are not yet known.  Concretely, we consider multiple black holes for which the horizons are disjoint unions of connected sums of $S^1 \times S^2$ or an $S^3$.  Note that  an analogue of Theorem \ref{theorem1} for multiple black holes has not been proven; however it has been conjectured that such a result should hold \cite{HHI}.  We remark that our restriction $n= n' = 0$  rules out the possibility that $H$ is a lens space other than $S^3$~\cite{Burton} (although it does not rule out connected sums of certain lens spaces) and we will not consider them here.  Note that asymptotically flat black lens metrics have been constructed (see \cite{Lens1,Lens2}) and although such solutions suffer from singularities, it seems reasonable to expect they could be removed by appropriate deformations.  It would be interesting to extend our results to such solutions.

A natural question is whether the topology of $H$ uniquely determines the topology of spatial sections of the domain of outer communication $\Sigma$.  A similar question arises from the observation  that Theorem \ref{theorem1} does not specify a unique smooth submanifold $B$ of $\mathbb{R}^4$ determined by $H$,  and different possibilities could yield different simply connected $\Sigma$s.  In the case of a black hole with horizon $H \cong S^3$, where $\cong$ means homeomorphism, there is  a unique possibility for $B$ and $\Sigma$, namely $B \cong B^4$ and $\Sigma \cong [0,\infty) \times S^3$ by the Schoenflies theorem \cite{Brown} in the topological category.  If $H \cong S^1 \times S^2$ or a connected sum of  $n$ copies of $S^1\times S^2$, we define the \emph{standard} black hole region as $B \cong S^1 \times B^3$ (where $B^n$ is the closed $n-$ball) or a connected union of $n$ copies of  $S^1 \times B^3$ respectively. It is unclear to us whether these are the unique possibilities, but we include a computation that shows that any other possibility is homologically, equivalent to the standard one.  In particular it will follow that the homology of $\Sigma$ only depends on $H$. We also perform some computations of higher dimensional homotopy groups of the domain of outer communication for the standard  doubly spinning black ring and black saturn.  We believe that our analysis could provide insights  in the context of proving existence of deformations of this initial data, as was performed for the Kerr black hole \cite{Dain} and for studying geometric inequalities on such manifolds \cite{Hollands2011}.

This note is organized as follows.  In Section 2 we consider three explicit examples of $(\Sigma, h_{ab},K_{ab})$ corresponding to the maximal slices of the Myers-Perry black hole, the doubly spinning black ring, and the Black Saturn solution.  In these cases $\Sigma$ is a simply connected asymptotically flat 4-manifold with inner boundary, and we examine some of its geometric and some homological properties. We show how the slice metric can be extended smoothly across the minimal surface corresponding to the horizon and compute their ADM mass.  In Section 3 we describe some general homological computations, show that the horizon determines the homological invariants of the slice for the class of horizons we are considering, and perform some  computations of higher homotopy groups. 

\section{Maximal slices of known black holes }  Consider a  five dimensional black hole spacetime $(M,g)$  with isometry group  $G = \mathbb{R} \times U(1)^2$. The vacuum Einstein equations imply orthogonal transitivity of $G$ \cite{GenWeyl} and we can therefore express the metric in the form
\begin{equation}
g =g_{\mu\nu}\d x^{\mu}\d x^{\nu}+g_{\alpha\beta}\d \xi^{\alpha}\d \xi^{\beta}.\label{1}
\end{equation}
where $\xi^{\alpha}=(t,\phi^i)$, $i=1,2$, are coordinates adapted to the Killing vector fields $\partial / \partial \xi^\alpha$ and $x^\mu$ are coordinates on the two-dimensional orbit space $M / G$. The metric components are independent of $\xi^\alpha$.   All explicitly known solutions are of this form.  We consider a spacelike hypersurface  $\Sigma$ of $M$ with normal $\d t$, induced metric $h$ and second fundamental form $K$.  The hypersurface  $(\Sigma,h_{ab},K_{ab})$ (indices $(a,b)$ refer to abstract coordinates on $\Sigma$) is a maximal slice of the spacetime if it has `$t-\phi^i$-symmetry' \cite{FMR}.  As pointed out in \cite{FMR}, this situation arises if (i) the Killing vectors associated to the axisymmetries are $\Phi^{a}_i=(\frac{\partial}{\partial \phi^i})^{a}$ (and hence extend to isometries of $(\Sigma,h_{ab},K_{ab})$), and (ii) $\phi^i\to -\phi^i$ is a diffeomorphism which preserves $h_{ab}$ but reverse the sign of $K_{ab}$. This symmetry implies
\begin{equation}
K_{ab}=2J^i_{(a}\Phi_{b)i}\,,\label{2}
\end{equation}
where $K_{ab}=\frac{1}{2}\mathcal{L}_{n}g_{mn}e^{m}_ae^{n}_b$, where $e^{m}_a$ are vierbeins used to pullback to $\Sigma$  and $n$ is a unit timeline normal vector to $\Sigma$.  For the class of metrics \eqref{1}, 
\begin{gather}\label{Jeq}
J^i_a=\frac{1}{2}\left(\partial_a n^{i}+\frac{g_{i j}g_{t j}-g_{j j} g_{t i}}{g^2_{i j}-g_{jj}g_{ii}}\partial_a n^{t}\right)\quad\quad i\neq j \quad\text{and}\quad i,j=1,2.
\end{gather}
Let $(H,\tilde{h}_{AB},\tilde{K}_{AB})$ denote the intersection of an event horizon in a stationary black hole spacetime of the form \eqref{1} with $\Sigma$.  This closed three-manifold is a marginally outermost trapped surface in the maximal slice  $(\Sigma,h_{ab},K_{ab})$ \cite{Gall}.  If $\tilde{s}$ is a unit outward normal to $S$ in $\Sigma$ then the condition that $S$ is an apparent horizon (marginally outer trapped) is \cite{FMR}  
\begin{equation}
\tilde{K}+\left(h^{ab}-\tilde{s}^a\tilde{s}^b\right)K_{ab}=0.\label{3}
\end{equation}
where $\tilde{K}=\text{div} \, \tilde{s} $ is the mean curvature of $S$. Since $S$ is invariant under the isometries generated by the $\Phi^i$, we will have $\tilde{s}^{a}\Phi^i_a=0$. Thus, as is well known,  the marginally outer trapped surface is also a minimal surface in $\Sigma$ . In fact, it has been shown that it is also a stable minimal surface \cite{FMR}.

The maximal slice $\Sigma$ is a complete, asymptotically flat Riemannian manifold.  For a single black hole there will be two asymptotic ends.  We will consider the complete Riemannian manifold with boundary by considering the part of $\Sigma$ from $S$ to one asymptotic end. Two important topological invariants of smooth four manifolds are the Euler number and signature. These can be computed from integrals over the curvature using the Gauss-Bonnet theorem and Hirzebruch signature theorem \cite{GH}. 
\normalsize In the majority of cases we are studying, $(\Sigma,h_{ab}, K_{ab})$ are sufficiently complicated and explicitly performing these integrals is too difficult.  It is easier to directly compute them from knowledge of the topological space.  By applying the Mayer-Vietoris theorem, we can calculate the relevant homology groups $\mathcal{H}_n(\Sigma)$. The Euler number is
\begin{equation}
\chi=\sum_{n=0}^{4} (-1)^{n}\text{dim} \mathcal{H}_n(\Sigma).
\end{equation} The calculation of the signature requires knowledge of the dimensions of the self and anti-self dual harmonic forms of $\Sigma$.  We will not pursue this here. \\
\subsection*{Singly-spinning Myers-Perry black hole}
The first case we will discuss is the singly-spinning Myers-Perry black hole solution in five dimensions with metric \cite{MP} 
\begin{gather}
ds^2=-\frac{\rho^2 \Delta^2}{\Sigma^2}\d t^2+\frac{\Sigma^2\sin^2\theta}{\rho^2}(\d\phi-\Omega \d t)^2+\frac{\rho^2 }{\Delta^2}\d r^2+\rho^2 \d\theta^2+r^2\cos^2\theta\d\psi.\\
\nonumber\\
\Delta^2=r^2+a^2-M^{2},\quad\quad\quad \rho^2=r^2+a^2\cos^2\theta,\\
\Sigma^2=(r^2+a^2)^2-a^2\Delta^2\sin^2\theta,\quad\ \Omega=\frac{aM^{2}}{\Sigma^2},
\end{gather}
This metric is parameterized by the mass parameter $M>0$ and the rotation parameter $a$ which are related to the ADM mass and angular momentum respectively.  As is well known, the solution admits a generalization with two independent angular momenta but we will restrict attention to the `singly-spinning' case (the doubly-spinning case does not present any relevant additional features). When  $a^2<M^2$, the spacetime has an event horizon at the positive root $\Delta(r)$,  $r_+=\sqrt{M^2-a^2}$. This metric can be extended across the event horizon and has a curvature singularity at $r=0$. The metric induced on a spatial hypersurface $t =$constant is 
\begin{eqnarray}
h&=&\frac{\rho^2}{\Delta^2}\d r^2+\rho^2\d\theta^2+\frac{\Sigma^2\sin^2\theta}{\rho^2}\d\phi^2+R^2\cos^2\theta \d\psi^2
\end{eqnarray}
The surface is guaranteed to be a maximal slice (the extrinsic curvature is given explicitly in \cite{FMR}) and it is straightforward to show explicitly that the event horizon intersects the slice on a minimal surface. We choose unit normal vector $\tilde{s}=\frac{dr}{\sqrt{h^{rr}}}$, so the mean curvature is
{\small
\begin{eqnarray}
\tilde{K}&=&\frac{\Delta\left(r\partial_r\Sigma^2+2\Sigma^2\right)}{2r\rho\Sigma^2}.
\end{eqnarray}
}\normalsize
and thus $\tilde{K}|_{r=r_+}=0$. \\
We can easily calculate the ADM mass of $\Sigma$ \cite{Bartnik}
\begin{equation}
m =\frac{1}{16\pi}\lim_{S\rightarrow i^0}\int_S{h^{ab}(\partial_{b}h_{ac}-\partial_{c}h_{ab})n^{c}\d S},\label{ADM}
\end{equation}
where S is a topological $3$-sphere with outward pointing unit normal $\tilde{n}^{a}$ and volume element, to find $m = 3M^2\pi/8$. 

The metric $h$ can be extended through the horizon by defining a function
\begin{equation}
H(r)=\frac{\Delta^2}{r^2}=\frac{(r-r_+)(r-r_-)}{r^2},
\end{equation}
and the transformation  
\begin{equation}
u(r)=\int_{r_+}^r{\frac{\d r'}{\sqrt{H(r')}}}=\sqrt{r^2+a^2-M^2}\label{6}
\end{equation}
The function $u(r)^2$ maps $[r_+,\infty)$ to $[0,\infty)$ and the metric is invariant under  $u(r)\leftrightarrow -u(r)$ . Note that $u(r)^2$ is an smooth function on $(r_+,\infty)$ but not differentiable at $r_+$, and that the inverse of $u(r)$ is $r(u)=\sqrt{u^2+M^2-a^2}$ which is a smooth mapping of $(-\infty,\infty)$ to $[r_+,\infty)$. Thus we have a smooth metric which can be extended to $u<0$:
\begin{eqnarray}
h = \frac{\rho^2}{r(u)^2}\d u^2+\rho^2 \d\theta^2+\frac{\Sigma^2}{\rho^2}\sin^2\theta \d\phi^2+r(u)^2\cos^2\theta \d\psi^2 
\end{eqnarray}
As for a maximal slice of Kerr \cite{Dain} (see also \cite{brestlec}), we can introduce a stereographic coordinate via $u = \cot \varphi$ where $0 < \varphi < \pi$. The region $u >0$ is mapped into $0 < \varphi < \pi/2$ and under the isometry $u\to -u$, there is another asymptotically flat region $u<0$ mapped to $\pi/2 < \varphi < \pi$. The metric can then be seen to extend smoothly to a metric on $S^4 - \{0\}-\{\pi\}$ where $(\varphi,\theta,\psi,\phi)$ are standard coordinates on $S^4$.  The `equator' $\varphi = \pi/2$ corresponds to the minimal $S^3$. We will focus on the region $u\geq 0$.  All components of the metric are smooth, positive, and uniformly bounded from zero functions for $u\geq 0$. Hence $(\Sigma,h)$ may be considered as a Riemanniani manifold with boundary. The metric is uniformly equivalent to the Eucldeain metric on $\mathbb{R}^4 \setminus B_1(0)$
\begin{equation}
h=\d u^2+(1+u)^2\left(\d\theta^2+\sin^2\theta \d\phi^2+\cos^2\theta \d\psi^2\right)
\end{equation} which is asymptotically flat as $u \to \infty$. The metric induced on the boundary $u=0$, is a metric on $S^3$ corresponding to the minimal surface $S$.  It can be seen that the region removed $u<0$ corresponds to an open ball.  In analogy with the case for Kerr, the slice has topology $\Sigma_{BH}\cong\mathbb{R}^4-\interior(B^4)$ and it is homeomorphic to $ \mathbb{R}\times S^3$. The region $\Sigma_{BH}$ given by $u\geq 0$ can be described
topologically as 
\begin{equation}
\Sigma_{BH}\cong\mathbb{R}^4-\interior(B^4)\cong [0,\infty) \times S^3\cong B^4 \# \mathbb{R}^4.\label{BHMP}
\end{equation}
$B^4$ is a closed 4-ball and $\interior(B^4)$ denotes an open 4-ball. The equivalences in the equation follow by observing that connected sum with $\mathbb{R}^4$ is equivalent to removing a point or a closed 4-ball, and connected sum with $B^4$ is equivalent to removing an open 4-ball. \\
For the Myers Perry maximal slice, it is straightforward to calculate the Euler number and signature directly from curvature integrals and we find
\begin{equation}
\chi=0, \quad\quad\quad\quad\quad \tau=0.
\end{equation} 
Since the slice $\Sigma_{BH}$ and $S^3$ are homotopic from equation (\ref{BHMP}), it follows that
\begin{equation}
\mathcal{H}_{n}(\Sigma_{BH}) = \left\{ 
   \begin{array}{l l}
     \mathbb{Z} & \quad n=0,3\\
     0 & \quad \text{others}
   \end{array} \right.
   \quad\quad\quad
   \chi(\Sigma_{BH})=\sum_{n=0}^{4} (-1)^{n}\text{dim} \mathcal{H}_n(\Sigma_{BH})=0
\end{equation}
Since $\Sigma_{BH}$ is simply connected, the first fundamental group of $\Sigma_{BH}$ is zero, i.e. $\pi_1(\Sigma_{BH})= 0$. Also, by the Hurewicz theorem $\pi_2(\Sigma_{BH})= 0$, and $\pi_3(\Sigma_{BH})= \mathbb{Z}$ . Other homotopy groups can be found from the standard reference \cite{Hatcher}. 

\subsection*{Doubly spinning black ring}
The second case we will discuss here is the doubly spinning black ring \cite{PS} (following the parametrization given in \cite{CHT})
\begin{eqnarray}
\d s^2&=&-\frac{H(y,x)}{H(x,y)}\left(\d t-\omega_{\phi}\d\phi-\omega_{\psi}\d\psi\right)^2-\frac{F(x,y)}{H(y,x)}\d\psi^2-2\frac{J(x,y)}{H(y,x)}\d\phi \d\psi
+\frac{F(y,x)}{H(y,x)}\d\phi^2\nonumber\\
&+&\frac{2\chi^2H(x,y)}{(1-\mu\nu)^2(x-y)^2}\left[\frac{\d x^2}{G(x)}-\frac{\d y^2}{G(y)}\right].
\end{eqnarray}
where
{\small
\begin{eqnarray*}
\omega_{\psi}&=& \frac{2\chi(\mu+\nu)}{H(y,x)}\sqrt{\frac{(1+\mu)(1+\nu)}{(1-\mu)(1-\nu)}}(1+y)\left[1 +\mu+\nu-\mu\nu+2\mu\nu x(1-y)+\mu\nu(1-\mu-\nu-\mu\nu)x^2y\right]\\
\omega_{\phi}&=& \frac{2\chi(\mu+\nu)}{H(y,x)}\sqrt{\mu\nu(1-\mu^2)(1-\nu^2)}(1-x^2)y,\\
G(x)&= &(1- x^2)(1+\mu x)(1 +\nu x),\\
H(x,y)&=& 1+(\mu +\nu)^2-\mu^2\nu^2+2(\mu +\nu)(x+\mu\nu y)(1-\mu\nu yx)+\mu\nu\left[1-(\mu+\nu)^2-\mu^2\nu^2\right]x^2y^2\\
J(x,y)&=&\frac{2\chi^2(\mu +\nu)\sqrt{\mu\nu}(1-x^2)(1-y^2)}{(1-\mu\nu)^2(x-y)}\Bigg[1+(\mu+\nu)^2- \mu^2\nu^2-\mu\nu\left[1-(\mu+\nu)^2-\mu^2\nu^2\right]xy\nonumber\\
&+&2\mu\nu(\mu+\nu)(x + y)\Bigg]
\end{eqnarray*}
\begin{eqnarray*}
F(x,y)&=& \frac{2\chi^2}{(1-\mu\nu)^2(x-y)^2}\Bigg[G(x)(1-y^2)\Bigg\{(1+\mu\nu)\left[1-(\mu+\nu)^2-2\mu\nu-\mu^2\nu^2\right]+(\mu+\nu)\nonumber\\
&\times&(1-\mu^2-\nu^2-3\mu^2\nu^2)y\Bigg\}+G(y)\Bigg\{2(\mu +\nu)^2 + (\mu +\nu)(1 +\mu^2)(1+\nu^2)x\nonumber\\
&+& (1+ \mu\nu)\left[1-(\mu+\nu)^2-2\mu\nu-\mu^2\nu^2\right]x^2+(\mu +\nu)\left[1-(\mu +\nu)^2
-\mu^2\nu^2(3-2\mu\nu)\right]x^3\nonumber\\
&+&\mu\nu(1-\mu\nu)\left[1-(\mu+\nu)^2-\mu^2\nu^2\right]x^4\Bigg\}\Bigg]
\end{eqnarray*}
}\normalsize
the range of parameter are $-1\leq x\leq 1$ and $y_h<y<-1$ and $0\leq \phi,\psi\leq 2\pi$. This metric is smooth except at the root of $G(y)$ at $y_h=-\frac{1}{\mu}$, corresponding to an event horizon. Spatial infinity occurs as $x,y\rightarrow -1$. The metric induced on a slice of constant time $t$ is 
\begin{eqnarray}
h&=&\frac{2\chi^2H(x,y)}{(1-\mu\nu)^2(x-y)^2}\left[\frac{\d x^2}{G(x)}-\frac{\d y^2}{G(y)}\right]-\left(\frac{H(y,x)}{H(x,y)}\omega^2_{\psi}+\frac{F(x,y)}{H(y,x)}\right)\d\psi^2 \nonumber
\\ &+& \left(-\frac{H(y,x)}{H(x,y)}\omega^2_{\phi}+\frac{F(y,x)}{H(y,x)}\right)\d \phi^2
-2\left(\frac{H(y,x)}{H(x,y)}\omega_{\phi}\omega_{\psi}+\frac{J(x,y)}{H(y,x)}\right)\d\psi \d\phi
\end{eqnarray}
Since the second fundamental form $K_{ab}$ is axisymmetric, it is obvious $K=0$ . Therefore, $\Sigma_{BR}$ is maximal spatial slice of black ring. Now we check  the apparent horizon $S=S^1\times S^2$ is minimal surface in $\Sigma_{BR}$ . We choose  $\tilde{s}_a \d x^a=\frac{\d y}{\sqrt{h^{yy}}}$ as unit outward normal vector. It is easy to compute the mean curvature \eqref{3}
\begin{eqnarray}
\tilde{K}&=&\sqrt{G(y)}m(x,y)
\end{eqnarray}
where $m(x,y)$ is a smooth function at $y=y_h$. As the horizon is at $y=y_h$, $\tilde{K}$ vanishes there. \\
The mass of the black ring spacetime can be computed directly form a Komar integral associated with the timelike Killing field $\partial/\partial t$. Here we demonstrate how it is found directly from the initial data using \eqref{ADM}. For simplicity we will consider the singly spinning black ring with zero angular momenta along the $S^2$, obtained by
\begin{equation}
\nu=0,\quad\quad\quad\quad R=\frac{2\chi^2(1-\mu)^2}{1-\lambda},
\end{equation}
We introduce a new  transformation $(x,y)\rightarrow(\rho,\theta)$ as following
\begin{gather}
R^2\rightarrow \tilde{R}^2\frac{1-\mu}{1-\lambda},\quad\quad x\rightarrow \frac{2\tilde{R}^2\cos^2\theta}{\rho^2}-1\quad\quad y\rightarrow -\frac{2\tilde{R}^2\sin^2\theta}{\rho^2}-1,
\end{gather}
then as $\rho \to \infty$, 
{\small
\begin{eqnarray}
h&\sim& \delta + \tilde{R}^2\left(\frac{1-3\mu}{\mu-1}\right)\left[\frac{\d\rho^2}{\rho^2}+\cos^4\theta \d\phi^2+\cos^2\theta\sin^2\theta \d\psi^2\right]+\frac{2\tilde{R}^2\lambda\cos^2\theta}{1-\lambda}\d\theta^2 \\
&+&\tilde{R}^2\frac{4\lambda\mu+2(1-3\nu)}{(\lambda-1)(\mu-1)}\cos^2\theta \frac{\d\rho^2}{\rho^2}
+\tilde{R}^2\frac{(\lambda+1)+\mu(\lambda-3)}{(\lambda-1)(\mu-1)}\sin^2\theta \d\psi^2,
\end{eqnarray}
} where $\delta$ is the flat Euclidean metric. Transforming to standard Cartesian coordinates 
\begin{equation}
(x,y,z,w) = (\rho\sin\theta \cos\phi,\rho\sin\theta \sin\phi,\rho\cos\theta \cos\psi,\rho\cos\theta \cos\psi),
\end{equation} we can straightforwardly compute
\begin{equation}
m=\frac{1}{16\pi}\int_{S^3_{\infty}}{h^{ab}(\partial_{b}h_{ac}-\partial_{c}h_{ab})n^{c}\d S}=\frac{3\pi R^2\mu}{2(1-\mu)(1+\mu^2)}.
\end{equation}
Returning to the general doubly-spinning case, we can  use a similar transformation as we performed for the Myers-Perry black hole to extend the metric through the horizon:
\begin{equation}
H(y)=\frac{G(y)}{h(y)}\quad\quad\quad\quad\quad h(y)=\frac{\mu(y_h+1)^2(y-1)(1+\nu y)}{4(1+y)^2},
\end{equation}
Then 
\begin{equation}
u(y)=\int_{y_h}^y{\frac{\d y}{\sqrt{H(y)}}}=\sqrt{\frac{y_h-y}{1+y}}
\end{equation}
The function $u(y)^2$ maps $[y_h,-1)$ to $[0,\infty)$ and it is an smooth function on $(y_h,-1)$ but not differentiable at $y_h$. The metric is invariant under the symmetry $u(y)\leftrightarrow -u(y)$ and hence can be extended beyond to $u<0$.   In addition, the inverse of $u(y)$ is $y(u)=\frac{y_h-u^2}{1+u^2}$, and $y(u)$ is a smooth function which maps $(-\infty,\infty)$ to $[y_h,-1)$. We find
\begin{equation}
h(u)=\frac{\mu}{4}\left(y_h-1-2u^2\right)\left(1+\nu y_h+(1-\nu)u^2\right).
\end{equation}
and we can write the local geometry of the slice as 
\begin{eqnarray}\label{slicering}
h&=&\left[-\frac{2\chi^2H(x,u)}{(1-\mu\nu)^2(x-y(u))^2h(u)}\right]\d u^2 -2\left(\frac{H(u,x)}{H(x,u)}\omega_{\phi}\omega_{\psi}+\frac{J(x,u)}{H(u,x)}\right)\d\psi \d\phi\nonumber \\ &+&\left[\frac{2\chi^2H(x,u)}{(1-\mu\nu)^2(x-y(u))^2(1+\mu x)(1+\nu x)(1+u)^2}\right](1+u)^2 \frac{\d x^2}{1-x^2}\nonumber\\
&+&\left[-\frac{H(u,x)^2\omega^2_{\psi}+F(x,u)H(x,u)}{H(x,u)H(u,x)(1+u)^2}\right](1+u)^2 \d\psi^2 \nonumber \\ &+&
\left[\frac{n(x,u)}{H(x,u)H(u,x)(1+u)^2}\right](1+u)^2(1-x^2)\d\phi^2
,
\end{eqnarray}
which $F(u,x)H(x,u)-H(u,x)^2\omega^2_{\phi}=(1-x^2)n(x,u)$ and $n(x,u)$ is smooth function.  Despite its fairly complicated appearance,  all functions of the above local metric are smooth, positive, and uniformly bounded away from zero  on  $\Sigma$. In particular, $(\Sigma,h)$ is complete Riemannian manifold with boundary corresponding to the apparent horizon of the black ring ($u=0$ in these coordinates). One can of course extend this metric to another isometric asymptotic region but we will not consider that here.   The metric \eqref{slicering} can be shown to be uniformly equivalent to a metric on $\Sigma$ of the form
\begin{equation}
h=\d u^2+(1+u)^2\left(\frac{\d x^2}{1-x^2}+(1-x^2)\d\phi^2+\d\psi^2\right),
\end{equation} and the local metric induced on the boundary minimal surface $u=0$ extends to a metric on $S^1\times S^2$. \par
The doubly spinning ring admits an extreme limit when $\nu=\mu$. In this case, the geometry has two asymptotic regions: in addition to the asymptotically flat end ($u \to \infty$) there is an asymptotically `cylindrical end' \cite{Dain} as $u \to 0$ which approaches the horizon. Thus $(\Sigma,h)$ is a complete Riemannian manifold without boundary and lies inside the black hole exterior.  In particular, setting $s=-\log u$, the extreme case of the local metric above admits a well-defined limit as $s\to \infty$: 
\begin{eqnarray}
h_{ext} &=& \left(\frac{2\chi^2 \mu^2 V(x)}{(1+\mu x)^2(1-\mu^2)}\right)\left[\frac{4 \d s^2}{1-\mu^2} + \frac{\d x^2}{(1-x^2)(1+\mu x)^2}\right] \nonumber \\
&+& \frac{8\chi^2 \mu^2(1-x^2)}{(1-\mu^2)V(x)}\left[ \d\phi - \frac{4\mu + 1 + \mu^2}{2\mu} \d\psi \right]^2 + \frac{4\chi^2(1+\mu)^2 \d\psi^2}{(1-\mu)^2}
\end{eqnarray} where $V(x) = (1+\mu^2)(1+x^2) + 4\mu x$. By an appropriate shift of $\phi$ one easily sees that the horizon is simply the product of $S^1$ with $S^2$ with inhomogeneous metric. The limit effectively cuts off the asymptotically flat end.  Although it is not obvious,  the above geometry is globally isometric to the cylindrical end of a maximal `$t$=constant' slice of a certain extreme boosted Kerr string. This relation between extreme black rings and boosted black strings was explicitly demonstrated at the level of near-horizon geometries in \cite{KLR, KL1} and indeed the above geometry can be considered as near-horizon initial data.

The maximal spatial slice $\Sigma_{BR}$ of the doubly spinning black ring spacetime has topology 
\begin{equation*} \Sigma_{BR} \cong \mathbb{R}^4-\interior(R) \end{equation*} where $R$ is a  regular neighborhood of an embedded $S^1$ in $\R^4$.  The choice of the embedding of $S^1$ into $\R^4$ is not relevant due to the fact that any pair of `locally flat'  embeddings of $S^1$ in $\mathbb R^4$ (or $S^4$) differ by a homeomorphism of $\mathbb R^4$ (respectively $S^4$), for this topological result see~\cite{Barker, Gluck}.

\begin{figure}
\centering
\subfloat[{ We delete $S^1\times B^3$ from $S^4$}]{\label{fig:1(b)}
\begin{tikzpicture}[scale=0.6, every node/.style={scale=0.7}]
\draw[fill=gray!30!white](-4,-1)--(2,-1)--(3.2,0)node[right,black]{$\mathbb{R}^3$}--(4,.7)
--(-2,.7)--(-4,-1);
\draw[fill=white] (0,0) ellipse (20pt and 8pt);
\draw[black](.7,0)--(.7,3);
\draw[black](-.7,0)--(-.7,1.5)node[left,black]{$S^1\times B^3$}--(-.7,3);
\draw (0,3) ellipse (20pt and 5pt);
\draw[black](.7,-1)--(.7,-3);
\draw[black](-.7,-1)--(-.7,-3);
\draw (0,-3) ellipse (20pt and 5pt);
\draw[fill=black] (0,4.3) circle (2pt)node[above=2pt,above,black]{$S^1=\mathbb{R}\cup\{\infty\}$};
\draw[thick,black,->](0,0)--(0,4);
\draw[thick,black](0,-1)--(0,-4);
\end{tikzpicture}}
\hfill
\subfloat[{ $S^2\times D^2$ is space around $S^1\times B^3$ }]{\label{fig:1(b)}
\begin{tikzpicture}[scale=0.65, every node/.style={scale=0.7}]
\draw[black,fill=gray!30!white] (0cm,2cm) arc (90:270:2cm)--cyclenode[,yshift=1.1cm,xshift=-.8cm,black]{$S^2\times D^2$};;
\draw[fill=white] (.7,1.95) ellipse (20pt and 5pt);
\draw[fill=white] (.7,-1.95) ellipse (20pt and 5pt);
\draw[fill=white] (.7,0) ellipse (20pt and 5pt);
\draw[black,fill=gray!30!white] (14mm,20mm) arc (90:-90:20mm)--cycle node[yshift=1cm,xshift=-.8cm,black,rotate=90]{$S^1\times B^3$};;
\draw[dashed,gray](.7,0) ellipse (77pt and 10pt);
\draw[dashed,gray](.7,1) ellipse (69pt and 8pt);
\draw[dashed,gray](.7,-1) ellipse (69pt and 8pt);
\draw[dashed,gray] (45mm,0mm) arc (0:360:40mm)--cycle node[right,black]{$S^4$};
\end{tikzpicture}}
\hfill
\subfloat[{ $S^2\times D^2\#\mathbb{R}^4$}]{\label{fig:1(c)}
\begin{tikzpicture}[scale=0.6, every node/.style={scale=0.7}]
\draw[fill=gray!30!white] (0,0) ellipse (80pt and 40pt)node[above=40pt,black]{$S^2\times D^2 \#\mathbb{R}^4$};
\draw (-1,.3) .. controls (-.5,-.3)and(.5,-.3) .. (1,.3);
\draw[fill=white] (-.76,.1) .. controls (-.3,.4)and(.3,.4)..(.76,.1);
\draw[fill=white] (-.76,.1) .. controls (-.3,-.23)and(.3,-.23)..(.76,.1);
\draw[fill=gray!30!white,gray!30!white] (2.3,.5) .. controls (3.4,.7)and(4,.8) .. (4.5,1.5)--(4.5,0)..controls (4,.5)and(3.5,.2)..(2.3,-.2)--(2.3,.5);
\draw (2.3,.5) .. controls (3.4,.7)and(4,.8) .. (4.5,1.5);
\draw[dashed,fill=gray!20!white,below](4.5,.73) ellipse (3pt and 21.3pt);;
\draw (2.3,-.2) .. controls (3.5,.2)and(4,.5) .. (4.5,0);
\draw[dashed](2.3,0.17) ellipse (4pt and 10pt);;;
\end{tikzpicture}}
\caption{ The black ring slice as $(S^2\times D^2) \# \R^4$.
{ (a) shows a regular neighborhood $R\cong S^1\times B^3$ of $S^1=\{\text{w-axis}\}\cup\{\infty\}$ is deleted  from $S^4\cong\mathbb{R}^4\cup \{\infty\}$. (b) the space obtained is homeomorphic to  $S^2\times D^2$  (c) The black ring slice topology, $S^2\times D^2\#\mathbb{R}^4$.}}
\label{fig:black ring slice}
\label{fig:02}
\end{figure}
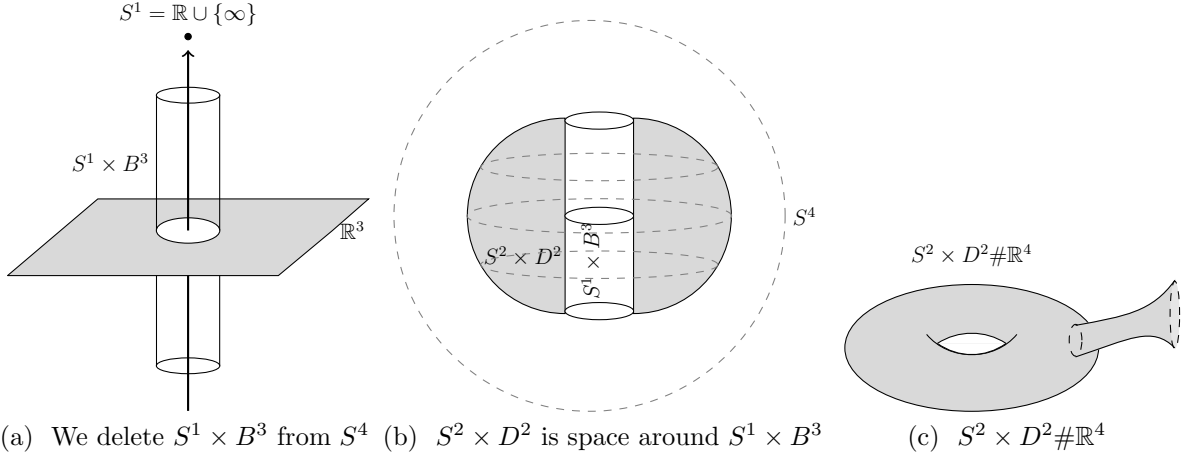

Consider the $4$-dimensional sphere $S^4$.  Let $B$ be a closed $4$-dimensional ball in $S^4$, let $R$ be a regular closed neighborhood of a locally flat embedded $S^1$ in  $S^4$, and assume that $B\cap R = \emptyset$. Since $S^4-B\cong \R^4$, it is immediate that
\begin{equation}\label{eq:sbr} \Sigma_{BR}\cong  S^4 -  \left[B \sqcup \interior(R)\right].\end{equation}
Regard $S^4$ as the one-point compactification $\mathbb{R}^4\cup\{\infty\}$ of $\mathbb{R}^4$, and without loss of generality assume that $R$ is a regular neighborhood of the $S^1$ formed by the w-axis of $\R^4$ together with $\{\infty\}$. Then one verifies that 
\begin{equation}\label{eq:s2d2} S^4-\interior(R) \cong S^2 \times D^2,\end{equation}
where $D^2$ denotes the $2$-dimensional closed disk; we refer the reader to figure~\eqref{fig:02}. It follows that 
\begin{equation}\label{eq:01}
\Sigma_{BR}\cong (S^2\times D^2)-B.
\end{equation}
Since removing a closed ball from a $4$-manifold is equivalent to a connected sum with $\R^4$, we also have that
\begin{equation*} 
\Sigma_{BR}\cong  (S^2\times D^2)\# \R^4, 
\end{equation*}
where $\#$ denotes the connected sum of manifolds. 
The homology of $\Sigma_{BR}$ is computed as follows. Since 
$\R^4 = \Sigma_{BR} \cup R$ and $R\cap \Sigma_{BR} = \partial R \cong S^1\times S^2$ and $R$ is homotopic to $S^1$;  the Mayer-Vietoris sequence for $\R^4 = \Sigma_{BR} \cup R$
\begin{equation*}  0\to  \mathcal{H}_i(S^1\times S^2) \to  \mathcal{H}_i(\Sigma_{BR}) \oplus \mathcal{H}_i(S^1)  \to 0, \quad i\geq 1.\end{equation*}
determines all homology groups of $\Sigma_{BR}$ and its Euler characteristic; namely
 \begin{equation*}
 \mathcal{H}_{n}(\Sigma_{BR}) = \left\{ 
    \begin{array}{l l}
      \mathbb{Z} & \quad n=0,2,3\\
      0 & \quad \text{others}
    \end{array} \right.
 \quad\quad\quad
 \chi(\Sigma_{BR})=\sum^{4}_{n=0}{(-1)^n \text{dim}\mathcal{H}_{n}(M)}=1.
 \end{equation*}
 This computation is a particular case of our Result \ref{result1}.

Since $\Sigma_{BR} \cong\R^4 - \interior(R)$ where $R$ is the regular neighborhood of an embedded $S^1$, an standard dimension argument shows that $\Sigma_{BR}$ is simply-connected. Since $\mathcal{H}_2(\Sigma_{BR})\cong \Z$,  the Hurewicz theorem shows that $\pi_2 (\Sigma_{BR}) \cong \Z$. More generally,  
\begin{equation}
\pi_1 (\Sigma_{BR})=0,\quad \quad \pi_2 (\Sigma_{BR})=\Z, \quad \quad \pi_3 (\Sigma_{BR})=\Z \oplus \Z \quad \quad \pi_4(\Sigma_{BR}) \neq 0,
\end{equation}
where  the claims about $\pi_3(\Sigma_{BR})$ and $\pi_4(\Sigma_{BR})$ are proved in Section 3 of the paper. 
The authors are not aware of an straight forward formal verification that  $\pi_3 (\Sigma_{BR})\cong \Z$.

\subsection*{Black Saturn}
The third case we will discuss is the Black Saturn vacuum solution \cite{EF2007}, which describes an asymptotically flat configuration representing the non-linear superposition of a singly-spinning black ring with a Myers-Perry black hole `inside'. The spacetime in the region exterior to both horizons written in the generalized Weyl coordinates $(\rho,z)$ takes the schematic form
\begin{equation}
\d s^2=-\frac{H_y}{H_x}\Bigg[\d t+\left(\frac{\omega_\psi}{H_y}+q\right)\d \psi\Bigg]^2+H_x\left[k^2P\left(\d\rho^2+\d z^2\right)+\frac{G_y}{H_y}\d \psi^2+\frac{G_x}{H_x}\d\phi^2\right]. 
\end{equation} where the various metric functions $H_i, G_i, \omega_\psi$ are functions of $(\rho,z)$ only and a general description of the coordinate system can be found in \cite{EF2007}.  There are five rod points in the structure of this metric and there are two horizons with topology $S^3$ and $S^1\times S^2$. These two horizon correspond to two intervals $z\in[a_1,a_2]$ and $z\in[a_3,a_4]$ on the axis $\rho=0$. The metric on a slice $t=$constant is 
\begin{eqnarray}
h&=&H_xk^2P\left(\d\rho^2+\d z^2\right)+G_x\d\phi^2+\left(-\frac{H_y}{H_x}\left(\frac{\omega_{\psi}}{H_y}+q\right)^2+\frac{H_xG_y}{H_y}\right)\d\psi^2.
\end{eqnarray}
Here we show the divergence of an outward-pointing normal to surfaces $\rho=0$ vanishes only on the intervals along the $z$ axis corresponding to the event horizons and hence they are marginally outer trapped surfaces in this spatial slice. As the Weyl coordinates degenerate at these points, we can introduce Kerr-like coordinates as given in \cite{CES} on each horizon interval. Let $a\in \mathbb{R}$ and $m>0$ such that
\begin{equation}
m^2-a^2=\left(\frac{a_j-a_i}{2}\right)^2.\label{weyl2}
\end{equation}
We define on $[a_i,a_j]=[a_5,a_4]$ or $[a_i,a_j]=[a_3,a_3]$ the following transformation
\begin{gather}
\rho=\sqrt{r^2-2mr+a^2}\sin\theta=\sqrt{(r-r_+)(r-r_-)}\sin\theta,\quad  z=\frac{a_i+a_j}{2}+(r-m)\cos\theta.\label{weyl1}
\end{gather}
where $r_{\pm}=m\pm\sqrt{m^2-a^2}$ with inverse
\begin{equation}
r=\frac{R_i+R_j}{2}+m, \quad\quad\quad\quad \cos\theta=\frac{R_i-R_j}{a_i-a_j}.\label{weyl3}
\end{equation}
and 
\begin{equation}
R_i=(r-r_+)+\frac{a_j-a_i}{2}\left(\cos\theta+1\right),\quad R_j=(r-r_+)+\frac{a_j-a_i}{2}\left(1-\cos\theta\right),
\end{equation}
Then we have
\begin{center}
     \begin{tabular}{ llllll}
     Half Axis & $\theta=\pi$ & $r\geq m+\sqrt{m^2-a^2}$ & $\Leftrightarrow$ & $\rho=0$ & $z\leq a_i$ \\ 
     Horizon & $0<\theta<\pi$ & $r=m\pm\sqrt{m^2-a^2}$ & $\Leftrightarrow$ & $\rho=0$ &  $a_i<z<a_j$ \\
     Half Axis& $\theta=0$  &  $r\geq m+\sqrt{m^2-a^2}$ & $\Leftrightarrow$ & $\rho=0$ &  $a_j\leq z$    \\
     \end{tabular}
\end{center}
In this coordinate the unit radial normal vector to horizon is $\tilde{s}_adx^a=\frac{dr}{\sqrt{h^{rr}}}$. In terms of the Weyl coordiantes,
\begin{equation}
\d r=\rho\frac{R_i+R_j}{2R_iR_j}\d\rho+\frac{R_j(z-a_i)+R_i(z-a_j)}{2R_iR_j}\d z=A \d\rho+B \d z.
\end{equation}
Also, in our metric $h_{\rho\rho}=h_{zz}$. Thus, the unit normal vector is 
\begin{equation}
\tilde{s}=\frac{A \d\rho}{\sqrt{h^{\rho\rho}(A^2+B^2)}}+\frac{B\d z}{\sqrt{h^{\rho\rho}(A^2+B^2)}}.
\end{equation} and computing the divergence gives the mean curvature
\begin{eqnarray}
\tilde{K}&=& f(z)\rho+O(\rho^2),\quad\quad\quad z\in[a_i,a_j],\\
\tilde{K}&=& g(z)+O(\rho),\quad\quad\quad z\in\mathbb{R}-\left([a_5,a_4]\cup [a_3,a_2]\right).
\end{eqnarray}
where $f(z)$ and $g(z)$ are smooth functions. Thus the horizons correspond to minimal surfaces in the slice,  and the mean curvature is nonzero on other intervals.  We can also extend the metric on the slice beyond each horizon (one also of course perform this extension in the full spacetime \cite{CES}). In the Kerr coordinates outlined in \cite{CES} the metric is
{\small
\begin{equation}
h=H_xk^2PR_iR_j\left(\frac{\d r^2}{(r-r_+)(r-r_-)}+\d\theta^2\right)+G_x \d\phi^2+\left(-\frac{H_y}{H_x}\left(\frac{\omega_{\psi}}{H_y}+q\right)^2+\frac{H_xG_y}{H_y}\right)\d\psi^2.
\end{equation}
}\normalsize
we define
\begin{equation}
H(r)=\frac{(r-r_+)(r-r_-)}{r^2},
\end{equation} 
then 
\begin{equation}
u(r)=\int_{r_+}^{r}{\frac{\d r}{\sqrt{H(r)}}}=\sqrt{r^2-2mr+a^2},
\end{equation}
The smooth function $u(r)^2$ maps $[r_+,\infty)$ to $[0,\infty)$ and it the metric is invariant under the transformation $u(r)\rightarrow -u(r)$. The inverse of $u(r)$ is $r(u)=m+\sqrt{u^2+m^2-a^2}$, which it is a smooth function that maps $(-\infty,\infty)$ to $[r_+,\infty)$.  As expected, we may consider the a maximal slice of the Black Saturn solution to be an asymptotically flat Riemannian manifold with two disconnected smooth boundary components corresponding to the minimal surfaces where the event horizon intersects the slice. If we extend to $u<0$ one enters an isometric region. A schematic illustration of a slice with multiple isometric regions is given in Figure 3. 
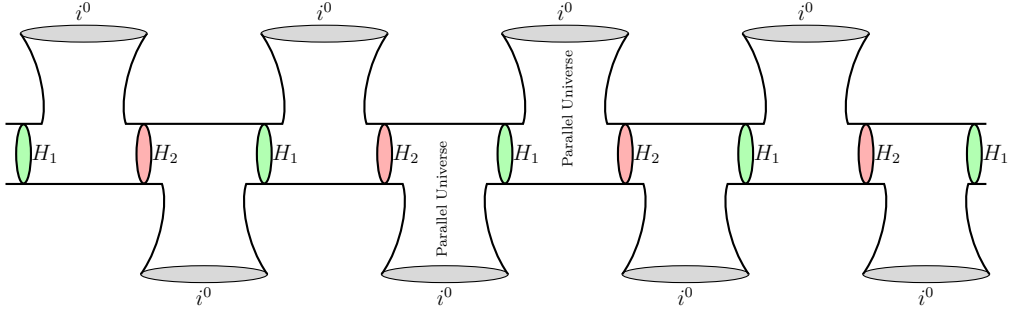
\begin{figure}
\centering
\begin{tikzpicture}[scale=0.8, every node/.style={scale=0.7}]
\draw[fill=gray!30!white] (2,-2) ellipse (30pt and 4pt)node[below=3pt,black]{$i^0$};
\draw[fill=gray!30!white] (-2,-2) ellipse (30pt and 4pt)node[below=3pt,black]{$i^0$};
\draw[fill=gray!30!white] (-6,-2) ellipse (30pt and 4pt)node[below=3pt,black]{$i^0$};
\draw[fill=gray!30!white] (-10,-2) ellipse (30pt and 4pt)node[below=3pt,black]{$i^0$};
\draw [thick] (3.05,-2).. controls (2.7,-1.5) and (2.6,-.8) .. (2.7,-.5).. controls (2.7,-.5)..(3,-.5) ;
\draw[thick] (-.95,-2).. controls (-1.3,-1.5) and (-1.4,-.8) .. (-1.3,-.5).. controls (-1.3,-.5)
      ..(1.3,-.5).. controls (1.4,-.8) and (1.3,-1.5)..(.95,-2) ;
\draw[thick] (-4.95,-2).. controls (-5.3,-1.5) and (-5.4,-.8) .. (-5.3,-.5).. controls (-5.3,-.5)
      ..(-2.7,-.5).. controls (-2.6,-.8) and (-2.7,-1.5)..(-3.05,-2) ;
\draw [thick](-8.95,-2).. controls (-9.3,-1.5) and (-9.4,-.8) .. (-9.3,-.5).. controls (-9.3,-.5)
      ..(-6.7,-.5).. controls (-6.6,-.8) and (-6.7,-1.5)..(-7.05,-2) ;
\draw[thick]  (-13.3,-.5)--(-10.7,-.5).. controls (-10.6,-.8) and (-10.7,-1.5)..(-11.05,-2) ;
\draw[fill=gray!30!white] (0,2) ellipse (30pt and 4pt)node[above=3pt,black]{$i^0$};
\draw[fill=gray!30!white] (-4,2) ellipse (30pt and 4pt)node[above=3pt,black]{$i^0$};
\draw[fill=gray!30!white] (-8,2) ellipse (30pt and 4pt)node[above=3pt,black]{$i^0$};
\draw[fill=gray!30!white] (-12,2) ellipse (30pt and 4pt)node[above=3pt,black]{$i^0$};
\draw [thick] (1.05,2).. controls (.7,1.5) and (.6,.8) ..(.7,.5).. controls (.7,.5)..(3,.5) ;
\draw[thick] (-2.95,2).. controls (-3.3,1.5) and (-3.4,.8) .. (-3.3,.5).. controls (-3.3,.5)
      ..(-.7,.5).. controls (-.6,.8) and (-.7,1.5)..(-1.05,2) ;
\draw[thick] (-6.95,2).. controls (-7.3,1.5) and (-7.4,.8) .. (-7.3,.5).. controls (-7.3,.5)
      ..(-4.7,.5).. controls (-4.6,.8) and (-4.7,1.5)..(-5.05,2) ;
\draw [thick](-10.95,2).. controls (-11.3,1.5) and (-11.4,.8) .. (-11.3,.5).. controls (-11.3,.5)
      ..(-8.7,.5).. controls (-8.6,.8) and (-8.7,1.5)..(-9.05,2) ;
\draw[thick]  (-13.3,.5)--(-12.7,.5).. controls (-12.6,.8) and (-12.7,1.5)..(-13.05,2) ;
\draw[thick,fill=green!30!white] (2.8,0) ellipse (3.5pt and 14pt)node[right,black]{$H_1$};

\draw[thick,fill=red!30!white] (1,0) ellipse (3.5pt and 14pt)node[right,black]{$H_2$};

\draw[thick,fill=green!30!white] (-1,0) ellipse (3.5pt and 14pt)node[right,black]{$H_1$};

\draw[thick,fill=red!30!white](-3,0) ellipse (3.5pt and 14pt)node[rotate=90,yshift=1.1cm,xshift=-.4cm,right,black,font=\scriptsize]{Parallel Universe}node[right,black]{$H_2$};

\draw[thick,fill=green!30!white] (-5,0) ellipse (3.5pt and 14pt)node[rotate=90,yshift=1.2cm,xshift=.4cm,left,black,font=\scriptsize]{Parallel Universe}node[right,black]{$H_1$};

\draw[thick,fill=red!30!white] (-7,0) ellipse (3.5pt and 14pt)node[right,black]{$H_2$};

\draw[thick,fill=green!30!white] (-9,0) ellipse (3.5pt and 14pt)node[right,black]{$H_1$};

\draw[thick,fill=red!30!white] (-11,0) ellipse (3.5pt and 14pt)node[right,black]{$H_2$};

\draw[thick,fill=green!30!white] (-13,0) ellipse (3.5pt and 14pt)node[right,black]{$H_1$};
\end{tikzpicture}
\caption{{\small The slice shape when there are two black holes with horizon $H_1$ and $H_2$.}}\label{fig:2}
\end{figure}
 
The topology of the black ring horizon and the Myers-Perry black hole horizon are  $S^1\times S^2$ and $S^3$, respectively. The topology of the spatial slice $\Sigma_{BS}$ of the black saturn is described as follows. Let $B^4$ be a closed $\epsilon$-ball in $\R^4$ and let $R$ be the regular closed neighborhood of a embedded $S^1$ in $\R^4$; in particular $R\cong S^1\times B^3$. Assume that $R\cap B^4 = \emptyset$. Then
\begin{equation*}  \Sigma_{BS}\cong\mathbb{R}^4-\left[\interior\left(B^4\right)\cup \interior(R)\right],\end{equation*}
or simply
\begin{equation*} 
\Sigma_{BS}\cong\mathbb{R}^4-\left\{\interior\left(B^4\right)\sqcup \left[S^1\times \interior\left(B^3\right)\right]\right\},\end{equation*}
 where $\sqcup$ denotes the disjoint union. Since $\R^4$ is homeomorphic to space obtained from $S^4$ by deliting a point, we have that 
\begin{equation*} \Sigma_{BS} = S^4-
\left\{ 
\{\text{point}\}  \sqcup \interior\left(B^4\right) \sqcup \left[S^1\times \interior\left(B^3\right) \right]
\right\}.
\end{equation*} 
Moreover, since $S^4-\left[S^1\times \interior\left(B^3\right)\right]$ is homemorphic to $S^2\times D^2$ we have
\begin{equation*} 
\Sigma_{BS}\cong \left( S^2\times D^2 \right) - \left[ \{\text{point}\} \sqcup \interior(B^4)\right].
\end{equation*}
Since deleting a point from a $4$-manifold is equivalent to removing a closed $4$-ball,  
\begin{equation}\label{eq:bss4} \Sigma_{BS} = S^4-
\left\{ 
B^4_1  \sqcup \interior\left(B^4_2\right) \sqcup \left[S^1\times \interior(B^3) \right]
\right\},
\end{equation} 
where $B^4_1$ and $B^4_2$ denote  disjoint closed $4$-balls in $S^4$; and also
\begin{equation*} \Sigma_{BS} = (S^2\times D^2)-
\left\{  \interior\left(B^4_2\right) \sqcup \left[S^1\times \interior\left(B^3\right) \right]
\right\}.
\end{equation*} 

Since removing a point from a $4$-manifold is equivalent to a connected sum with $\R^4$, and removing an open ball is equivalent to a connected sum with a closed ball,  we have
\begin{equation*}
\Sigma_{BS}\cong  \left( S^2\times D^2 \right) \#\mathbb{R}^4\#B^4.   
\end{equation*}
The homology groups of $\Sigma_{BS}$ can be computed directly using  the excision theorem; we perform a more general computation in the next section for which the next table is a particular case:
\begin{equation}
\mathcal{H}_n(\Sigma_{BS})  = \left\{ 
   \begin{array}{l l}
     \mathbb{Z} & \quad n=0,2\\
     \mathbb{Z}^2 & \quad n=3\\
     0 & \quad \text{others}
   \end{array} \right.
\quad\quad\quad
\chi(\Sigma_{BS})=\sum^{4}_{n=0}{(-1)^n \text{dim}\mathcal{H}_{n}(\Sigma_{BS})}=0,
\end{equation}
Since $\Sigma_{BS}$ is simply connected, the first fundamental group is zero, i.e. $\pi_1(\Sigma_{BS})= 0$. Also, by the Hurewicz theorem $\pi_2(\Sigma_{BS})= \mathbb{Z}$. In section 3 we shows the following: 
\[ \pi_1(\Sigma_{BS})= 0,\quad \quad \pi_2(\Sigma_{BS})= \Z, \quad \quad \pi_3(\Sigma_{BS})=\Z\oplus \Z \oplus \Z \quad \quad \pi_4(\Sigma_{BR}) \neq 0 \;.\] 
 Note that it should be possible to extend these computations to other known multiple black hole configurations, e.g. \cite{EF, ER2008, Iguchi} .

\section{Some topological aspects}  In this section we will first consider the computation of certain toplogical invariants for slices with standard black hole regions as defined previously.  Note that the geometrical condition of maximality is not used in the following. We define the standard region of a black  hole with $H \cong \# m\,(S^1\times S^2)$ as $\mathcal{N}_{\epsilon}\left(\vee_{i=1}^{l}S_{i}^1 \right)$ where $\mathcal{N}_{\epsilon}$ represents a regular neighbourhood.  This definition requires us to show that there are no knotted embeddings of $\mathcal{N}_{\epsilon}\left(\vee_{i=1}^{l}S_{i}^1 \right)$ into $\mathbb{R}^4$. We address this issue in the second part of this section. 
\subsection{Topological computations for standard black hole regions} 
Our first result computes the Euler number of a slice of a five-dimensional spacetime  containing $m$ black holes (the existence of which is consistent with all known constraints).  Observe that this means we are considering spacetimes which contain a disjoint union of horizons, each of which is consistent with the horizon classification of \cite{HHI}.  Although there are no explicit solutions for such geometries, we can still discuss aspects of their topology.

%

\begin{result} \label{result1} Consider an asymptotically flat stationary spacetime $(M,g)$ containing $m=n_1 + n_2 + n_3$ black holes and horizon
\\
\begin{equation}
H\cong\left(\coprod_{i=1}^{n_1}S^3\right)\coprod\left(\coprod_{i=1}^{n_2}\left(S^1\times S^2\right)\right)\coprod\left(\coprod_{i=1}^{n_3}\#l\,(S^1\times S^2)\right).
\end{equation} Assume that the domain of outer communication has the form $\mathbb{R} \times \Sigma$ where $\Sigma \cong \mathbb{R}^4 - B$ and $B$ is the standard black hole region for $H$.  Then the Euler number of $\Sigma$ is $\chi = 1-n_1+n_3(l-1)$ and the homology of $\Sigma$ is given by expression~\eqref{homologySigma}.
\end{result}
\begin{remark}
Note that the hypotheses of this result hold in the case when $(M,g)$ admits a $\mathbb{R} \times U(1)$ isometry and $H$ is connected i.e. there is a single black hole \cite{HHI}. 
\end{remark}

\begin{proof}[Proof of Result \ref{result1}] In this case the black hole region is  
\begin{equation}
B\cong\left(\coprod_{i=1}^{n_1}\interior(B^4)\right)\coprod\left(\coprod_{i=1}^{n_2}\left(S^1\times \interior(B^3)\right)\right)\coprod\left(\coprod_{i=1}^{n_3}\mathcal{N}_{\epsilon}\left(\vee_{i=1}^{l}S_{i}^1 \right)\right).
\end{equation}  
The homology of the black hole region is
\begin{equation}
\mathcal{H}_n(B)=\bigoplus_{i=1}^{n_1}\mathcal{H}_n\left(\interior(B^4)\right)\bigoplus_{i=1}^{n_2}\mathcal{H}_n\left(S^1\times \interior(B^3)\right)\bigoplus_{i=1}^{n_3}\mathcal{H}_n\left(\mathcal{N}_{\epsilon}\left(\vee_{i=1}^{l}S_{i}^1 \right)\right),
\end{equation}
Since $\mathcal{H}_n\left(\mathcal{N}_{\epsilon}\left(\vee_{i=1}^{l}S_{i}^1 \right)\right)=\bigoplus_{i=1}^{l}\mathcal{H}_n(S^1_{i})$ we have
\begin{equation}
\mathcal{H}_n(B)  = \left\{ 
   \begin{array}{l l}
     \mathbb{Z}^{m} & \quad n=0\\
     \mathbb{Z}^{n_2+n_3l} & \quad n=1\\
     0  & \quad \text{others}
   \end{array} \right.
\end{equation}
Also the homology of the horizon is 
\begin{equation}
\mathcal{H}_n(H)=\bigoplus_{i=1}^{n_1}\mathcal{H}_n\left(S^3\right)\bigoplus_{i=1}^{n_2}\mathcal{H}_n\left(S^1\times S^2\right)\bigoplus_{i=1}^{n_3}\mathcal{H}_n\left(\#l\,\left(S^1\times S^2\right)\right)
\end{equation}
By the long exact sequence and the excision theorem, a calculation shows
\begin{equation}
\mathcal{H}_n\left(\#l\,(S^1\times S^2)\right)  = \left\{ 
   \begin{array}{l l}
     \mathbb{Z} & \quad n=0,3\\
     \mathbb{Z}^{l} & \quad n=1,2\\
     0 & \quad  n \geq 4
   \end{array} \right.
\end{equation}
then
\begin{equation}
 \mathcal{H}_n(H)  = \left\{ 
    \begin{array}{l l}
      \mathbb{Z}^{m} & \quad n=0,3\\
      \mathbb{Z}^{n_2+n_3l} & \quad n=1,2\\
      0  & \quad n\geq 4
    \end{array} \right.
\end{equation} Since $\mathbb{R}^4\cong\Sigma\cup B$ with $B\cap\Sigma\cong H$, from the long exact sequence 
\begin{equation}\label{MVSigma}
\mathcal{H}_{n+1}(\mathbb{R}^4)\rightarrow \mathcal{H}_n (H) \rightarrow \mathcal{H}_n(B) \oplus  \mathcal{H}_n (\Sigma ) \rightarrow \mathcal{H}_{n}(\mathbb{R}^4)
\end{equation}
we deduce the following:
\begin{equation}
\begin{tabular}{ccccc}
\hline
 $\text{dim}\mathcal{H}_n$ & $\mathbb{R}^4$ & $H$&  $B$ & $\Sigma$\\
\hline
$n=0$  & 1 & $m$          & $m$          & 1 \\
$n=1$  & 0 & $n_2+n_3l$   & $n_2+n_3l$ & 0 \\
$n=2$  & 0 & $n_2+n_3l$   & 0            & $n_2+n_3l$ \\
$n=3$  & 0 & $m$          & 0            & $m$ \\
$n\geq4$& 0 & 0           & 0            & 0 \\
\hline 
\end{tabular} 
\end{equation}The homology of a slice of spacetime containing $m$ stationary black holes is
\begin{equation} \label{homologySigma}
 \mathcal{H}_n(\Sigma)  = \left\{ 
    \begin{array}{l l}
      \mathbb{Z} & \quad  n=0\\
      \mathbb{Z}^{n_2+n_3l} & \quad n=2\\
      \mathbb{Z}^{m} & \quad n=3\\
      0  & \quad  n=1\,\text{and}\, n\geq 4
    \end{array} \right.
\end{equation}
Then the Euler number is 
\begin{gather}
\chi(\Sigma)=\sum^{4}_{n=0}{(-1)^n \text{dim}\left(\mathcal{H}_n\left(\Sigma\right)\right)}=1-n_1+n_3\left(l-1\right).
\end{gather}
\end{proof}
We remark that under the assumptions of Result \ref{result1}, Hurewics' theorem implies $\pi_2(\Sigma)=\mathbb{Z}^{n_2+n_3l} $.  Further, suppose that  $(M,g)$ is a asymptotically flat stationary spacetime containing a black hole with horizon $H$ such that $\mathbb{R}^4\cong\Sigma\cup B,\, H\cong\Sigma\cap B$. We are unaware whether the following statement is true: topologically, the only possibilty for the black hole region $B$ with horizon $\#m\, (S^1\times S^2)$ is a regular neighbourhood of $\vee_{i=1}^mS_{i}^1$.  We can, however, show the following:

\begin{result}
Suppose $(M,g)$ is an asymptotically flat stationary spacetime containing a black hole with horizon $H$ such that $\mathbb{R}^4\cong\Sigma\cup B$ and $H\cong\Sigma\cap B$  where $H \cong \#m\,(S^1\times S^2)$. Then the homology of $B$ is the same as the homology of the standard black hole region for $H$. Moreover, the homology of $\Sigma$ is the same as the homology of the $\Sigma$ obtained by removing the standard black hole region from $\mathbb{R}^4$. 
\end{result}
\begin{proof} Consider the Mayer-Vietoris sequence 
\begin{equation}
...\mathcal{H}_{n+1}(\mathbb{R}^4)\rightarrow \mathcal{H}_n (\#m\,(S^1\times S^2)) \rightarrow \mathcal{H}_n(B) \oplus  \mathcal{H}_n (\Sigma ) \rightarrow \mathcal{H}_{n}(\mathbb{R}^4)... 
\end{equation} Observe that
\begin{itemize}
\item $\mathcal{H}_0(B)=  \mathcal{H}_0(\Sigma) = \mathbb{Z}$ since   $B$ and $\Sigma$ are  connected.
\item $\mathcal{H}_1(B)= \mathbb{Z}^m$. This follows since by topological censorship \cite{topcen}, asymptotic flatness implies $\Sigma$ is simply connected so $\mathcal{H}_1(\Sigma) = 0$.  
\item $\mathcal{H}_2 (B)= 0$ and $\mathcal{H}_2 (\Sigma )= \mathbb{Z}^m$. Indeed, by Alexander duality~\cite[Thm. 3.44]{Hatcher},  $\mathcal{H}_2 (\Sigma  \cup \{ \infty \}) = \mathcal{H}_2 (S^4-B) = \mathcal{H}^1(B) = \mathbb{Z}^m$.   Then, observe that, 
$\mathcal{H}_2 (\Sigma ) =  \mathcal{H}_2 (\Sigma  \cup \{ \infty \})$ since removing an interior point of a $4$-manifold does not change the second homology group. This last statement is well known and is proved as follows: consider the sequence for the pair $(M^4,M^4-\{p\})$ and use that $\mathcal{H}_2(M^4, M^4-\{p\}) = \mathcal{H}_2 (\mathbb{R}^4, \mathbb{R}^4-\{p\}) = \mathcal{H}_2 (S^3)$ which holds by excision. 
\item $\mathcal{H}_3(B)= 0$ and $\mathcal{H}_3 (\Sigma ) = \mathbb{Z}$.
Analogously, by Alexander duality~\cite[Thm. 3.44]{Hatcher},  $\mathcal{H}_3 (\Sigma  \cup \{ \infty \}) = \mathcal{H}_3 (S^4-B) = \widetilde {\mathcal{H}}^0(B) = \mathbb{Z}$. 
Then $\mathcal{H}_3 (\Sigma )= \mathcal{H}_3 (\Sigma  \cup \{ \infty \}) = \mathbb{Z}$.
\item $\mathcal{H}_n(B)=\mathcal{H}_n(\Sigma)= 0$ for $n \geq 4$ since $B$  and $\Sigma$ are 4-manifolds with boundary.
\end{itemize}
 Observe this agrees with the homology of the standard case for $H \cong \#m\,(S^1\times S^2)$. 
\end{proof}

\subsection{Standard Black Hole Regions are Well Defined}

Recall that we previously defined the \emph{standard} black hole region $B$ for a black hole with horizon a connected sum $\# m\, (S^1 \times S^2)$ of $m$ copies of $S^1\times S^2$ to be a smooth regular neighbourhood of a \emph{subspace} homemorphic to  $\vee_{i=1}^{m}S_{i}^1$ of $S^4$.  We want to make sure that generically different ways to consider $B$ are equivalent. Specifically, if $B_1$ and $B_2$  are two different possible standard regions, then there is  diffeomorphism $h\colon S^4 \to S^4$ such that $h(B_1)=B_2$.  In fact the stronger statement that $h$ is differentiable and isotopic to the identity map follows from standard results in differential topology as informally  described below.

First, the notion of subspace is restricted to being a subcomplex in a triangulation of $S^4$, and  smooth regular neighborhood is defined as done by Hirsch \cite{Hirsch-text}. Then any pair of smooth regular neighborhoods of a subcomplex $\vee_{i=1}^{m}S_{i}^1$ of $S^4$ are differentiable isotopic. By reasons of dimension, any pair of subcomplexes of $S^4$ homemorphic to $\vee_{i=1}^{m}S_{i}^1$ of $S^4$ are isotopic, and this  isotopy extends to an ambient differentiable isotopy by the Isotopy Extension Theorem~\cite{Hirsch}.

\subsection{Higher homotopy groups for maximal spatial slides}

The computations below rely on the excision theorem for homotopy groups:
\begin{thm}[Excision for Homotopy Groups]\cite[Thm 4.23]{Hatcher}
Let $X$ be a CW-complex decomposed as the union of subcomplexes $A$ and $B$ with nonempty connected intersection $C = A\cap B$. If $(A,C)$ is $m$-connected
and $(B,C)$ is $n$-connected, $m,n \geq 0$, then the map $\pi_i (A,C) \to \pi_i(X,B)$ induced by inclusion is an isomorphism for $i< m+n$ and a surjection for $i =m+ n$.
\end{thm}

\subsubsection*{The doubly spinning black ring maximal spatial slice}

In this part we verify that $\pi_3 (\Sigma_{BR}) = \Z$ and $\pi_4 (\Sigma_{BR}) \neq 0$.  From~\eqref{eq:sbr},
\begin{equation} \Sigma_{BR}\cong  S^4 -  \left(B^4 \cup \interior(R)\right),\end{equation}
where  $B^4$ is a closed $4$-dimensional ball in $S^4$,   $R \cong S^1\times B^3$ is a regular closed neighborhood of a locally flat embedded $S^1$ in $S^4$, and $B^4\cap R = \emptyset$. 
 Consider
\begin{equation} S^4 = M \cup N\end{equation} 
where 
\begin{equation} M= S^4-\interior(R),\quad \text{ } \quad  N= S^4- \interior (B),\quad \text{ and } \quad  \overline  \Sigma_{BR}  = M\cap N \end{equation}  
Observe that $\Sigma_{BR}$ and its closure in $S^4$ are  homotopy equivalent. The following connectivity properties hold:
\begin{itemize}
\item {\it Claim 1.} The pair $\left (N,  \overline \Sigma_{BR}  \right)$ is $2$-connected.
\item {\it Claim 2.} The pair $\left (M,  \overline \Sigma_{BR}  \right) $ is $3$-connected.
\end{itemize}
First we verify that $\pi_3 (\Sigma_{BR}) = \Z$ and $\pi_4 (\Sigma_{BR}) \neq 0$ assuming that  both claims hold, and after the computation we will verify claims.   The excision theorem for homopy groups stated above~\cite[Thm 4.23]{Hatcher} applied to $S^4$ as the union of $M$ and $N$ together with claims 1 and 2 imply that the map induced by inclusion
\begin{equation}\label{eq:06} \pi_4 \left (M,  \overline \Sigma_{BR}  \right)   \overset\cong\longrightarrow \pi_4 (S^4, N) 
\end{equation}
is an isomorphism. 
The long exact sequence of homotopy groups for the pair $(S^4, N)$ and the fact that $N$ is contractible shows that 
\begin{equation} \label{eq:07} 0=\pi_{i+1} (N) \to \pi_{i+1} (S^4) \overset\cong\to \pi_{i+1} (S^4, N) \to \pi_i (N)=0. \end{equation}
The isomorphisms of~\eqref{eq:06} and ~\eqref{eq:07} together with the fact $\pi_4 (S^4) \cong \Z$ yield 
\begin{equation} \pi_4 \left (M,  \overline \Sigma_{BR}  \right)  \cong \Z.\end{equation}
The long exact homotopy sequence for the pair $(M, \overline \Sigma_{BR})$ yields 
\begin{equation}  \label{(M,Sigma)}
\cdots \to \pi_4 (M) \to \pi_4(M,\overline{\Sigma}_{BR})\to \pi_3 (\overline{\Sigma}_{BR}) \to \pi_3 (M) \to \pi_3 (M,\overline{\Sigma}_{BR})\to \cdots .
\end{equation} 
Observe that the second claim implies that $\pi_3 (M,\overline{\Sigma}_{BR})=0$.  In~\eqref{eq:s2d2} we show that $M\cong S^2\times D^2$ and hence $\pi_i (M) \cong \pi_i (S^2)$.  Since $\pi_4 (S^2)\cong \Z_2$ and $\pi_3 (S^2)\cong \Z$ and $\pi_4 \left (M,  \overline \Sigma_{BR}  \right)  \cong \Z$, then~\eqref{(M,Sigma)} yields a short exact sequence 
\begin{equation} \Z_2 \to \Z \to \pi_3( \overline{\Sigma}_{BR}) \to \Z \to 0 .\end{equation}
The first map on the left must be trivial. Then, since $\pi_3( \overline{\Sigma}_{BR})$ is an abelian group, it follows that \begin{equation} \pi_3(  {\Sigma}_{BR}) \cong \pi_3( \overline{\Sigma}_{BR}) \cong \Z\oplus \Z.\end{equation}
 Similarly, we have the long exact sequence of homotopy groups
\begin{equation}
\cdots \to \pi_4 (\overline{\Sigma}_{BR}) \to \pi_4(M) \to \pi_4(M,\overline{\Sigma}_{BR}) \to \cdots
\end{equation} Using the well known results $\pi_5(S^4) = \Z_2$, $\pi_4(S^2) = \Z_2$, and $\pi_4(S^4)= \Z$,  the sequence above implies that there is a surjective map 
$ \pi_4 (\overline{\Sigma}_{BR}) \to \Z_2 $ and hence $\pi_4(\overline{\Sigma}_{BR})$ is non-trivial.  \\
\par \noindent \emph{Verification of claim 1.}  Since $\Sigma_{BR}$ is $1$-connected and 
$N$ is contractible, the long exact sequence for 
$(N, \overline\Sigma_{BR}   )$ yields 
\begin{equation} 0=\pi_{i+1} (N) \to \pi_{i+1} \left( N,  \overline\Sigma_{BR} \right) \to \pi_i \left(  \overline\Sigma_{BR} \right) \to \pi_i (N)=0.\end{equation}
It follows that $\left( N,  \overline\Sigma_{BR} \right)$ is $2$-connected as claimed. \\
\par \noindent \emph{Verification of claim 2.}  Observe that the pair $(B, \partial B)$ is $2$-connected.  Indeed, since $B$ is a $4$-dimensional ball, the long exact sequence of homotopy groups for $(B, \partial B)$,
\begin{equation} 0=\pi_{i+1} B \to \pi_{i+1} (B, \partial B) \to \pi_i (\partial B) \to \pi_i B=0,\end{equation}
shows that \begin{equation}\label{eq:02}\pi_{i+1}(B, \partial B) \cong \pi_i (\partial B) \cong \pi_i( S^3).\end{equation}
Analogously, observe that the pair $(\overline \Sigma_{BR}, \partial B)$ is $1$-connected. Indeed, since $\partial B\cong S^3$,  the long exact sequence for $(\overline \Sigma_{BR}, \partial B)$ shows that
\begin{equation} 0=\pi_1 ( S^3) \to \pi_1 (\overline \Sigma_{BR}) \to \pi_1 (\overline \Sigma_{BR}, \partial B) \to \pi_0 (S^3)=0. \end{equation}
Since $\pi_1 (\Sigma_{BR} ) \cong \pi_1 (\overline \Sigma_{BR}) = 0$, it follows that $\pi_1 (\overline \Sigma_{BR}, \partial B)$ is trivial.  Since $(B, \partial B)$ is $2$-connected and $(\overline \Sigma_{BR}, \partial B)$ is $1$-connected, a direct application of the excision theorem for higher homotopy groups~\cite[Thm 4.23]{Hatcher} applied to $M=\overline \Sigma_{BR} \cup B$ shows that  the composition
\begin{equation}\label{eq:03}  \pi_i (B, \partial B) \overset\cong\to \pi_i (M, \overline \Sigma_{BR}), \quad i=1,2 \end{equation}
is an isomorphism, and the composition 
\begin{equation}\label{eq:04} \pi_3 (B, \partial B) \to \pi_3 (M, \overline \Sigma_{BR}) \end{equation}
is surjective.  Then equations~\eqref{eq:02},~\eqref{eq:03} and~\eqref{eq:04} yield that $\pi_i (M, \overline \Sigma_{BR})$ is trivial for $0\leq i\leq 3$, hence $(M, \overline \Sigma_{BR})$ is $3$-connected as claimed.

\subsubsection*{The Black Saturn maximal spatial slice}

We now verify that $\pi_3 (\Sigma_{BS}) = \Z^3$ and $\pi_4 (\Sigma_{BS})$ is not trivial. The computation follows the same strategy as the previous computation. From~\eqref{eq:bss4}, 
\begin{equation*} 
\Sigma_{BS} = S^4- \left\{  B^4_1  \sqcup \interior\left(B^4_2\right) \sqcup  \interior(R)  \right\}
\end{equation*}
where   $B_1^4$ and $B_2^4$ are disjoint closed $\epsilon$-balls in $S^4$,  $R\cong S^1\times B^3$ is a regular closed neighborhood of an embedded $S^1$ in $S^4$ and  $R\cap (B_1^4\cup B_2^4) = \emptyset$. Consider 
\begin{equation*} 
 S^4 = M\cup N
\end{equation*}
where  
$M = S^4- \interior(R) \cong S^2\times D^2$  and 
$N = S^4- \interior(B_1 \cup B_2) \cong S^3\times \R$  and 
$M\cap N = \overline \Sigma_{BS}$. 
The following connectivity properties hold:
\begin{itemize}
\item {\it Claim 1.} The pair $\left (N,  \overline \Sigma_{BS}  \right)$ is $2$-connected.
\item {\it Claim 2.} The pair $\left (M,  \overline \Sigma_{BR}  \right) $ is $3$-connected.
\end{itemize}
First we verify that $\pi_3 (\Sigma_{BS}) = \Z^3$ assuming that  both claims hold, and after the computation we will verify the claims.  By the excision theorem for homotopy groups applied to $S^4 = M\cup N$ we obtain that the map induced by inclusion
\begin{equation} \label{eq:bs11} 
\pi_i(M, \overline \Sigma_{BS}) \to \pi_i  (S^4, N) 
\end{equation}
is an isomorphism for $i\leq 4$. 
Consider the long exact sequence for $(S^4, N)$ (note that $N$ is homotopy equivalent to $S^3$)
\begin{equation}  
\label{eq:bsi}\cdots \to \pi_4(S^3)\to \pi_4(S^4) \to \pi_4 (S^4, N) \to \pi_{3}(S^3)\to \pi_3(S^4) \to \cdots 
\end{equation}
Since $\pi_4(S^3)\cong \Z_2$ is finite, $\pi_4(S^4)\cong \Z$  and $\pi_3 (S^4)=0$ we have a short exact sequence 
\begin{equation} 
0\to \Z \to \pi_4 (S^4, N) \to \Z \to 0.
\end{equation}
Since  $\pi_4 (S^4, N)$ is an abelian group, it follows that
\begin{equation} \label{eq:bs13} 
\pi_4 (S^4, N) \cong \Z\oplus \Z.
\end{equation}  
Now the long exact sequence for the pair $(M,\overline{\Sigma}_{BR})$  is
\begin{equation} \label{eq:bsii}
\cdots \to \pi_4(M) \to \pi_4(M,\overline{\Sigma}_{BR}) \to \pi_3(\overline{\Sigma}_{BR}) \to \pi_3(M) \to \pi_3(M,\overline{\Sigma}_{BR}) \to \cdots
\end{equation}  Recall $M$ is homotopy equivalent to $S^2$, giving $\pi_4(M) = \Z_2, \pi_3(M) = \Z$. 
As a consequence of the isomorphism \eqref{eq:bs11} and \eqref{eq:bs13}, $\pi_4(M,\overline{\Sigma}_{BR}) \cong \Z \oplus \Z$. 
A further application of the sequence \eqref{eq:bsi} gives $\pi_3(M,\overline{\Sigma}_{BR}) \cong 0$. Inserting these results into \eqref{eq:bsii} yields a short exact sequence
\begin{equation}
0\to \Z \oplus \Z \to \pi_3( \overline{\Sigma}_{BR} )\to \Z \to 0
\end{equation}  
from which it follows $\pi_3(\overline{\Sigma}_{BR}) = \Z^3$. 
Finally, to demonstrate $\pi_4(  \overline{\Sigma}_{BR})$ is non trivial, we again consider the following portion of the long exact sequence for the pair $(M,\overline{\Sigma}_{BR})$:
\begin{equation*}
\ldots \to \pi_4(\overline{\Sigma}_{BR}) \to \pi_4(M) \to \pi_4(M,\overline{\Sigma}_{BR}) \to \ldots
\end{equation*}  
Upon inserting $\pi_4(M) = \Z$ and our previous result $\pi_4(M,\overline{\Sigma}_{BR}) = \Z \oplus \Z$ one finds
\begin{equation*}
\ldots \to \pi_4(\overline{\Sigma}_{BR}) \to \Z_2 \to \Z \oplus \Z \to \ldots
\end{equation*}
 Clearly, the map $\Z_2 \to \Z^2$ must be trivial, implying that the map $\pi_4(\overline{\Sigma}_{BR}) \to \Z_2$  is a surjection. This shows that $\pi_4(\overline{\Sigma}_{BR})$ is non-trivial. \\
\par \noindent \emph{Verification of claim 1.}  Since $N$ is homotopy equivalent to $S^3$,  the long exact sequence for $\left (N,  \overline \Sigma_{BS}  \right)$ provides an exact sequence 
\begin{equation} \pi_{i+1}(S^3) \to \pi_{i+1}(S^3\times \R, \overline \Sigma_{BS}) \to \pi_i(\Sigma_{BS}) \to \pi_i (S^3). \end{equation}
It follows that 
\begin{equation} \pi_{i+1}(S^3\times \R, \overline \Sigma_{BS}) \cong \pi_i(\Sigma_{BS}), \quad i=0,1.\end{equation} Since $\Sigma_{BS}$ is $1$-connected, it follows that $(N, \overline \Sigma_{BS})$ is $2$-connected. \\
\par \noindent \emph{Verification of claim 2.} 
 Consider $S^2\times D^2$ as the union of $\overline \Sigma_{BS}$ and $P=B_1 \cup B_2^4\cup \gamma$ where $\gamma$ is a simple path in $\overline \Sigma_{BS}$ from $\partial B_1^4$ to $\partial B_2^4$.  Let $\partial P$ denote the intersection of $P$ and $\overline \Sigma_{BS}$.
The introduction of $\gamma$ is to guarantee that $\partial P$ is  connected; indeed observe that  $\partial P$ is homotopic to a wedge of a pair of $3$-spheres $S^3\vee S^3$.

First we show $(P, \partial P)$ is $3$-connected.  Since $P$ is contractible, the long exact sequence for $(P, \partial P)$ implies that
\begin{equation}\label{eq:bs30}  \pi_{i+1}(P, \partial P) \cong \pi_i (S^3 \vee S^3) \quad i\geq 0 .\end{equation}
Since $S^3$ is $3$-connected, the wedge $S^3\vee S^3$ is $2$-connected by the main result in~\cite{Hilton55}. Then~\eqref{eq:bs30} implies that $(P, \partial P)$ is $3$-connected.

Now we show that $(\overline \Sigma_{BS}, \partial P)$ is $1$-connected. The long exact sequence for this pair provides the exact sequence (of sets)
\begin{equation} 0=\pi_1 (\partial P) \to \pi_1(\overline \Sigma_{BS}) \overset\cong \to \pi_1(\overline \Sigma_{BS}, \partial P) \to \pi_0 (\partial P)=0  \end{equation}
Since $\overline \Sigma_{BS}$ is $1$-connected, it follows that 
$(\overline \Sigma_{BS}, \partial P)$ is $1$-connected.

Then the excision theorem for homotopy groups applied to $M=\overline \Sigma_{BS} \cup P$ implies that 
\begin{equation} \pi_i(P, \partial P) \cong \pi_i (M, \Sigma_{BS}), \quad 0\leq i\leq 3.\end{equation}
Since $(P, \partial P)$ is $3$-connected, we have that $(M, \Sigma_{BS})$ is $3$-connected

\section*{Acknowledgements}
 AA is supported by a graduate scholarship from Memorial University. HKK and EMP are each supported by an NSERC Discovery Grant. We would like to thank Tom Baird and Juan Souto for useful comments.  HKK especially thanks Kristin Schleich and Don Witt for useful discussions and clarifications.

\end{document}